\documentclass[12pt]{article}

\begin{document}


\newcommand{\bb}{\begin{equation}}
\newcommand{\ee}{\end{equation}}
\newcommand{\bbb}{\begin{eqnarray}}
\newcommand{\eee}{\end{eqnarray}}
\newcommand{\vc}[1]{\mbox{$\vec{{\bf #1}}$}}
\newcommand{\mc}[1]{\mathcal{#1}}
\newcommand{\del}{\partial}
\newcommand{\lk}{\left}
\newcommand{\ave}[1]{\mbox{$\langle{#1}\rangle$}}
\newcommand{\re}{\right}
\newcommand{\pd}[1]{\frac{\del}{\del #1}}
\newcommand{\pdd}[2]{\frac{\del^2}{\del #1 \del #2}}
\newcommand{\Dd}[1]{\frac{d}{d #1}}
\newcommand{\sech}{\mbox{sech}}
\newcommand{\pref}[1]{(\ref{#1})}

\newcommand
{\sect}[1]{\vspace{20pt}{\LARGE}\noindent
{\bf #1:}}
\newcommand
{\subsect}[1]{\vspace{20pt}\hspace*{10pt}{\Large{$\bullet$}}\mbox{ }
{\bf #1}}
\newcommand
{\subsubsect}[1]{\hspace*{20pt}{\large{$\bullet$}}\mbox{ }
{\bf #1}}

\def\ie{{\it i.e.}}
\def\eg{{\it e.g.}}
\def\cf{{\it c.f.}}
\def\etal{{\it et.al.}}
\def\etc{{\it etc.}}

\def\AA{{\cal A}}
\def\BB{{\cal B}}
\def\CC{{\cal C}}
\def\DD{{\cal D}}
\def\EE{{\cal E}}
\def\FF{{\cal F}}
\def\GG{{\cal G}}
\def\HH{{\cal H}}
\def\II{{\cal I}}
\def\JJ{{\cal J}}
\def\KK{{\cal K}}
\def\LL{{\cal L}}
\def\MM{{\cal M}}
\def\NN{{\cal N}}
\def\OO{{\cal O}}
\def\PP{{\cal P}}
\def\QQ{{\cal Q}}
\def\RR{{\cal R}}
\def\SS{{\cal S}}
\def\TT{{\cal T}}
\def\UU{{\cal U}}
\def\VV{{\cal V}}
\def\WW{{\cal W}}
\def\XX{{\cal X}}
\def\YY{{\cal Y}}
\def\ZZ{{\cal Z}}

\def\sinh{{\rm sinh}}
\def\cosh{{\rm cosh}}
\def\tanh{{\rm tanh}}
\def\sgn{{\rm sgn}}
\def\det{{\rm det}}
\def\exp{{\rm exp}}
\def\sh{{\rm sh}}
\def\ch{{\rm ch}}

\def\ell{{\it l}}
\def\str{{\it str}}
\def\lp{\ell_{{\rm pl}}}
\def\ls{\ell_{{\str}}}
\def\gs{g_\str}
\def\gym{g_{Y}}
\def\r11{R_{11}}

\input{epsf}

\begin{titlepage}
\rightline{EFI-98-53}

\rightline{hep-th/9810224}

\vskip 3cm
\centerline{\Large{\bf Black Holes and the SYM Phase Diagram. II}}

\vskip 2cm
\centerline{
Emil Martinec\footnote{\texttt{ejm@theory.uchicago.edu}}~~ and ~~
Vatche Sahakian\footnote{\texttt{isaak@theory.uchicago.edu}}}
\vskip 12pt
\centerline{\sl Enrico Fermi Inst. and Dept. of Physics}
\centerline{\sl University of Chicago}
\centerline{\sl 5640 S. Ellis Ave., Chicago, IL 60637, USA}

\vskip 2cm

\begin{abstract}
The complete phase diagram of objects in M-theory compactified on
tori $T^p$, $p=1,2,3$, is elaborated.  Phase transitions occur 
when the object localizes on cycle(s) (the Gregory-Laflamme transition),
or when the area of the localized part of the horizon
becomes one in string units (the Horowitz-Polchinski correspondence point).  
The low-energy, near-horizon geometry that governs a given phase
can match onto a variety of asymptotic regimes.
The analysis makes it clear that the matrix conjecture is a special case
of the Maldacena conjecture.
\end{abstract}

\end{titlepage}

\newpage
\setcounter{page}{1}

\section{Introduction and summary}
The Matrix and Maldacena conjectures~\cite{MAT1,MAT2,MALDA1} 
boldly propose a great
simplification of the dynamics of M-theory in certain limits.
Both postulate that all of M-theory -- restricted to a sector
of particular boundary conditions -- is equivalent to 
super Yang-Mills theory (SYM).  Conversely, we learn that
nonperturbative super Yang-Mills is of a complexity equivalent
to M-theory.  Taking both conjectures into account, we expect
that maximally supersymmetric Yang-Mills on $T^p$, $p\le3$, has
dynamics of perturbative gauge theory, of near-horizon
D-brane geometry, and of light-cone (LC) M-theory,
all as different regimes in some grand phase diagram.

The aim of this paper is to map out the thermodynamic phase diagram 
of this theory. The finite temperature vacuum will acquire 
various geometrical realizations, 
and a rich thermodynamical structure will emerge.
One conclusion is that, taking as inputs
the Maldacena conjecture and the various duality symmetries of M-theory
established at the level of the low energy dynamics, the Matrix conjecture
is a necessary output.
The point is that one can glue a given near-horizon structure
onto several descriptions of the asymptotic region
at large distance from the object.  The choice of asymptotic
description fixes a duality frame, and hence the physical interpretation
of the parameters.  Nevertheless, one can have \eg\ a near-extremal
D3-brane in type IIB theory whose thermodynamics is described
by the 11D Schwarzschild black hole geometry.
Which is the appropriate low-energy geometry to describe
the horizon physics is governed by the proper size of the torus
near the horizon, which depends on the horizon radius or entropy
of the object.  Qualitatively, at high entropy the Dp brane 
perspective is appropriate, as the horizon torus is large;
at low entropies, the horizon torus is small, and a T-dualized
D0 brane description is appropriate.  This is the T-duality
used in Matrix theory.
To avoid cumbersome changes of notation, we will phrase
the discussion in the language of light cone M-theory,\footnote{For instance,
the charge quantum $N$ will be referred to as longitudinal momentum
rather than D-brane charge.}
even when the entropy is large enough that a dual D-brane description
is appropriate; we hope this does not cause undue confusion.

The statement of the Maldacena conjecture identifies a dual geometrical
structure with the (possibly finite temperature)
vacuum of a SYM Quantum Field Theory (QFT). Physics of
elementary excitations off this
vacuum are mapped onto the physics of elementary probes in the background
of the near-horizon geometries of branes~\cite{GUBSER,WITHOLO}. 
A non-trivial UV-IR correspondence relates
radial extent in the bulk geometry and
energy scale of such excitations~\cite{SUSSHOLO,PEETPOLCH}. 
On the other hand, equilibrium thermodynamic states of
SYM excitations get associated to geometries with thermodynamic character,
\ie\ supergravity vacua with finite area horizons.
The UV-IR correspondence then amounts to essentially the
holographic relation between area in the bulk and entropy in the SYM.
Alternatively, it relates temperature in
a QFT to radial extent in a non-extremal bulk~\cite{PEETPOLCH}.

When studying the thermodynamics of 
M-theory via that of a SYM, we then
encounter on the phase diagram patches with dual supergravity realizations.
Geometrical considerations will identify stability domains, transition
curves and various equations of state. The global structure of the
diagram should reflect the myriad of duality symmetries that M-theory
is endowed with; geometry will paint the global structure of
M-theory's finite temperature vacuum. The basic idea then becomes
a systematic analysis of various supergravity solutions; the underlying
strategy goes as follows:
A $10$D or lower-dimensional near-extremal supergravity 
solution must satisfy the following restrictions:
\begin{itemize}
\item The dilaton at the horizon must be small. Otherwise, in a IIA
theory, we need to lift to an $11$D M-theory; in a IIB theory, we need
to go to the S-dual geometry. 
This amounts to a change of description --
a reshuffling of the dominant degrees of freedom -- 
without any change in the equation of state. 
\item The curvature at the horizon must be smaller than the string scale.
Otherwise, the dynamics of massive string modes becomes relevant. By
the Horowitz-Polchinski correspondence principle~\cite{CORR1}, a string theory
description emerges -- an excited string, 
or a perturbative SYM gas reflecting
weakly coupled D-brane dynamics. This is generally associated with a change
of the equation of state; in the thermodynamic limit, we may expect 
critical behavior associated with a phase transition.
This criterion can easily be estimated for various cases
when one realizes that the 
curvature scale is set by the horizon area divided by cycle sizes 
measured at the horizon; \ie\ the localized horizon area should be
greater than order one in string units.\footnote{In general,
the horizon will be localized in some dimensions and delocalized (stretched) 
in others.  The area of the `localized part of the horizon'
means the area along the dimensions in which the horizon is localized.}
\item Cycles of tori on which the geometry may be wrapped, as measured at
the horizon, must be greater than the string scale~\cite{RABIN}. 
Otherwise, light winding
modes become relevant and the T-dual vacuum describes the proper
physics~\cite{GIVEON}.  
We expect no critical behavior in the thermodynamic limit,
since the duality is merely a change of description.
\item The horizon size of the geometry must be smaller than the torus 
cycles as measured at the horizon~\cite{LAFLAMME1,LAFLAMME2}. 
Otherwise, the vacuum smeared
on the cycles is entropically favored. We expect this to be associated
with a phase transition, one due to finite size effects, and it is
associated generally with a change of the equation of state. However, it
is also possible that there is no such entropically favored transition
by virtue of the symmetry structure of a particular
smeared geometry, so we expect no change of phase.  Intuitively,
a system would only localize itself in a more symmetrical solution 
to minimize free energy. 
\end{itemize}

\noindent
On the other hand,
given an $11$D supergravity vacuum, 
a somewhat different set of restrictions applies:
\begin{itemize}
\item The curvature near the horizon must be smaller than the Planck scale.
By the criterion outlined above, we see that, for unsmeared geometries,
this is simply the statement that $S> 1$; \ie\ quantum gravity effects are
relevant for low entropies. For large enough longitudinal momentum $N$, 
this region of the phase diagram is well away from the region
of interest.
\item The size of cycles of the torus as measured at the horizon must be greater
than the Planck scale. Otherwise, we need to go to the IIA solution descending
from dimensional reduction on a small cycle. We expect no change of 
equation of state or critical behavior.
\item The size of the M-theory cycles, including the 
light cone longitudinal box,
as measured at the horizon, must be bigger than the horizon size. Otherwise,
the geometry gets smeared along the small cycles. This is expected to be
a phase transition due to finite size physics.
\end{itemize}

\noindent
Applying these criteria, we then select
the near extremal geometry dual to a SYM theory on the
torus, and navigate the phase diagram via duality transformations
suggested by the various restrictions.
We will then be mapping out the phase
diagram of M-theory, or alternatively SYM QFT; 
we now present the results of such an analysis.

\begin{figure}[p]
\epsfxsize=13cm \centerline{\leavevmode \epsfbox{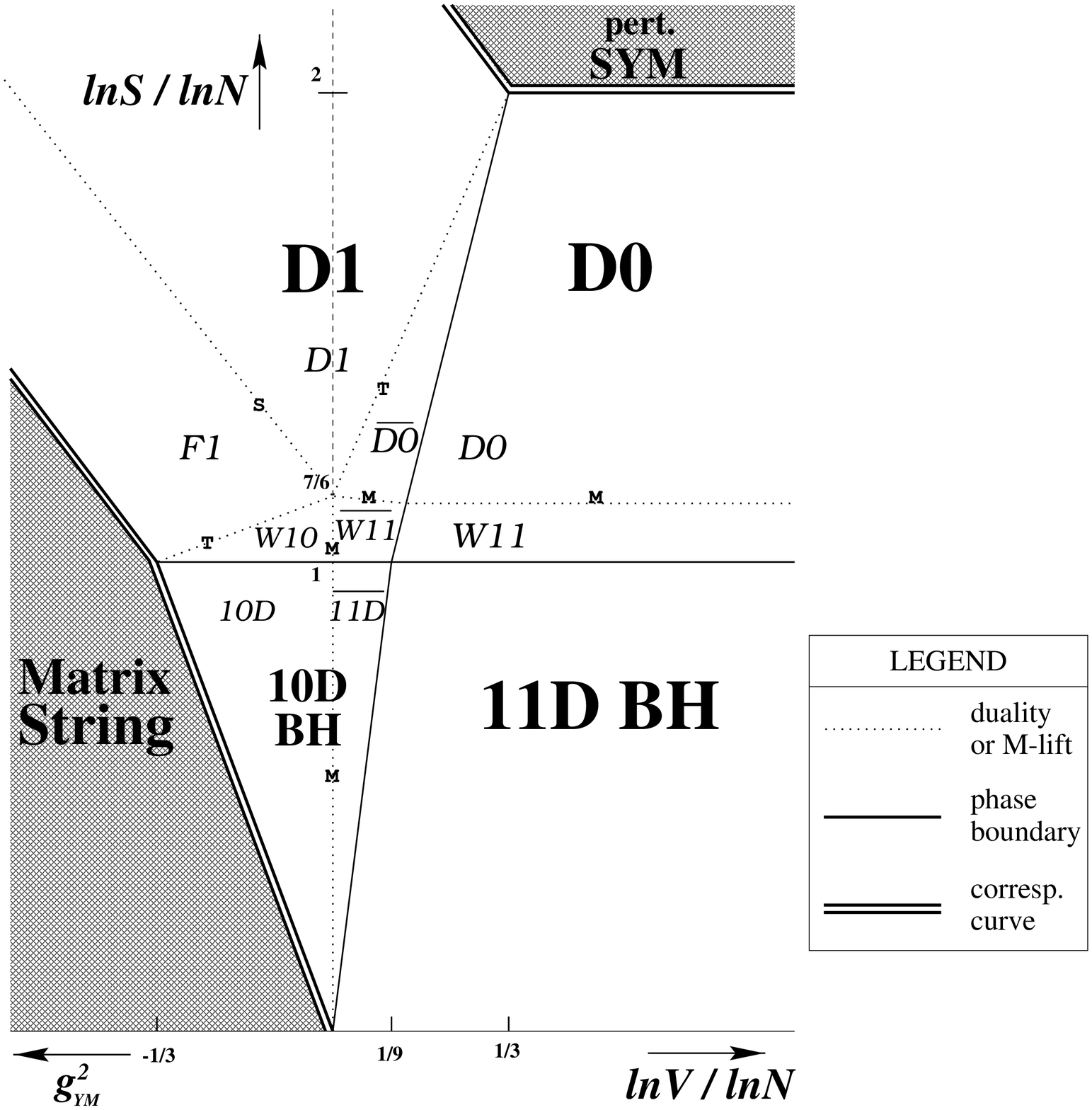}}
\caption{\sl Phase diagram of Light Cone M theory on $T^1$;
$S$ is entropy, $V$ is the radius of the circle in Planck units, $N$ is
the longitudinal momentum. The geometry label dictionary is as follows:
D0: black D0; 
$\overline{D0}$: black D0 smeared on $V$;
D1: black D1; 
F1: black IIB string; 
W10: black IIA wave; 
W11: $11$D black wave; 
$\overline{W11}$: $11$D black wave smeared on $V$; 
$10$D BH: IIA LC black hole;
$11$D BH: LC M-theory black hole; 
$\overline{11D}$ BH: LC M-theory BH smeared on $V$. 
$M$, $T$ and $S$ stand for respectively an M-duality
(such as reduction, lift or M flip on $T^3$), a T-duality 
curve, and an S duality transition.}
\label{phase1}
\end{figure}

Figure~\pref{phase1} depicts the thermodynamic phase diagram of 
light cone M-theory
on the circle; the axes are entropy $S$, 
and the radius of the circle measured in Planck units $V$; 
$N$, the longitudinal momentum charge, is fixed. 
In general, the effective SYM coupling and torus radii are
\bb\label{geff}
\Sigma=\frac{\lp^2}{\r11 V}\quad,\qquad g_{\rm eff}^2\sim 
\gym^2 N T^{p-3} \sim V^{-p} N \lk(\frac{T \lp^2}{\r11}\re)^{p-3}
\ee
($T\sim E/S$ is the temperature). 
On the diagram, the behavior of the effective SYM coupling depends on the
equation of state governing a given region under consideration.
For $p<4$, the effective coupling increases as we move vertically 
downward in the D0 phase, and diagonally toward the bottom-left in the
Dp brane phase. The raw Yang-Mills coupling 
(measured at the scale of the torus) increases horizontally
as we move toward smaller M theory volumes $V^p$ for all $p$.
The unshaded areas
are described by various supergravity solutions, while the shaded
regions do not have dual geometrical descriptions. 
On the upper right, there is a perturbative
$1+1$D SYM gas phase living on the circle;
and a Matrix string phase
in the IR of the SYM, characterized by the emergence of $Z_N$ order,
on the left at strong coupling.
Dotted lines denote various duality transformations on the supergravity 
solutions; the equation of state is unchanged upon crossing
such a line, since duality is merely a change of description.  
The solid lines denote localization transitions; 
double solid lines are curves associated
with the principle of correspondence of Horowitz-Polchinski. 
These lines demarcate phase boundaries, 
where the equation of state changes.
In total, we have six different thermodynamic phases. 
We observe the emergence of a self-duality
point at $S\sim N^{7/6}$, $V\sim 1$. 
That the various patches do not overlap
is a self-consistency check
on the logical structure of the picture. 
For high entropies, localization effects are circumvented; 
the phases are the ones studied in~\cite{MALDA2}.
The triple point on the upper right corner was the one studied 
in~\cite{RABIN}.
The lower left triple point was the one studied in~\cite{LMS}. This picture
patches together these previous results on one diagram, in addition
to identifying one additional triple point and the self-duality point.
The oblique correspondence curve in the upper right corner can easily
be seen to correspond to the point where the effective dimensionless 
Yang-Mills coupling is of order one.  More interesting is the
horizontal correspondence curve along $S\sim N^2$ 
starting at $V\sim N^{1/3}$. 
From the perturbative SYM side, 
it is where the thermal wavelength becomes of order
the size of the box dual to $V$; 
from the D0 phase side, it is a Horowitz-Polchinski correspondence curve. 
As we will see below, the temperature jumps discontinuously
as function of the entropy across this line.
In~\cite{RABIN}, a separate phase consisting of a gas of super Quantum
Mechanics excitations was identified with this curve when the phase diagram
is plotted on the temperature-'t Hooft coupling plane.
This transition may be associated with rich microscopic physics.
From the thermodynamic perspective, as the
transition is crossed, dynamics is transferred from
local excitations in $p$+1 SYM to that of its zero modes; 
and Dp brane charge of the perturbative SYM is
traded for longitudinal momentum charge of light cone M-theory.
This process is one of several paths on the phase diagram
relating the Maldacena and Matrix conjectures
\footnote{Ideas relating the Matrix and Maldacena conjectures were also
discussed in~\cite{JOHN1,JOHN2,JOHN3}.}.

\begin{figure}[p]
\epsfxsize=13cm \centerline{\leavevmode \epsfbox{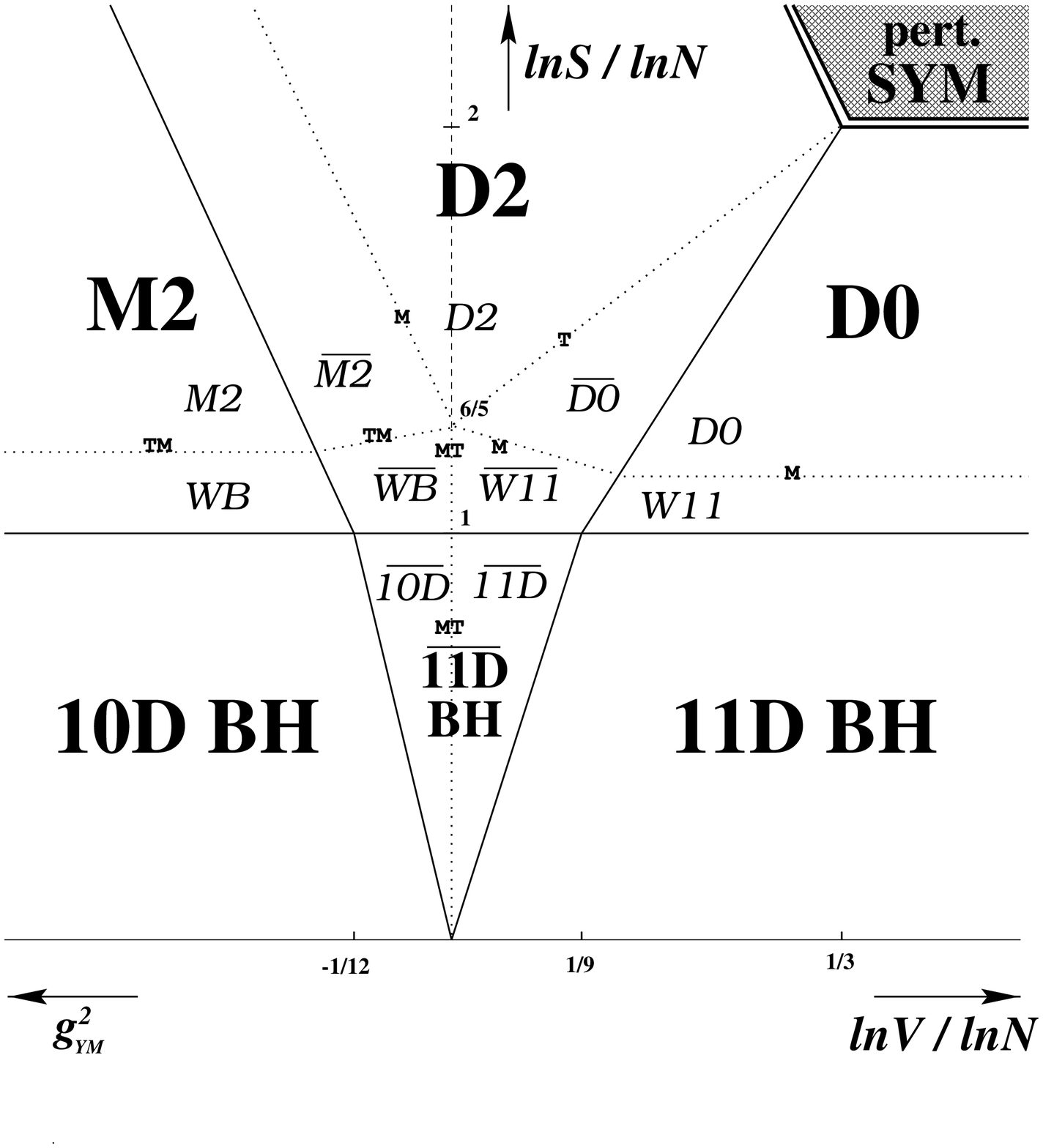}}
\caption{\sl Phase diagram of Light Cone M theory on $T^2$;
$V$ is the radius of the circle in Planck units.
The geometry label dictionary is as follows:
D0: black D0; 
$\overline{D0}$: black D0 smeared on $V$;
D2: black D2; 
M2: black membrane; 
$\overline{M2}$: black membrane smeared on a dual circle;
WB: black IIB wave; 
$\overline{WB}$: black IIB wave smeared on a dual circle;
W11: $11$D black wave; 
$\overline{W11}$: $11$D black wave smeared on $V$; 
$11$D BH: light cone M-theory black hole; 
$\overline{11D}$ BH: light cone M-theory black hole smeared on $V$;
$10$D BH: IIB light cone black hole; 
$\overline{10D}$ BH: IIB light cone black hole smeared on a dual circle.
}
\label{phase2}
\end{figure}

\begin{figure}[p]
\epsfxsize=14cm \centerline{\leavevmode \epsfbox{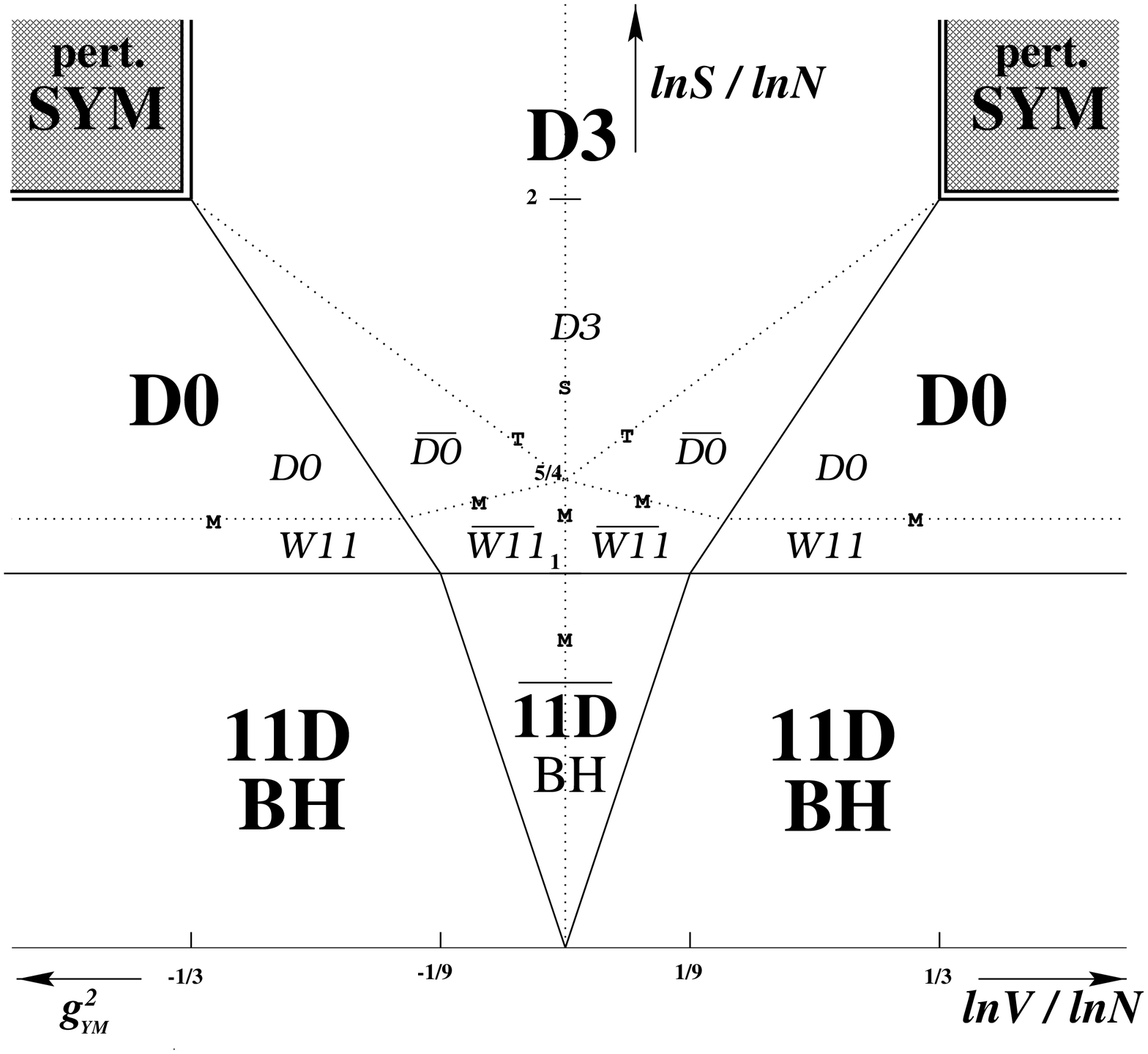}}
\caption{\sl Phase diagram of Light Cone M theory on $T^3$.
D0: black D0; 
$\overline{D0}$: black D0 smeared on $V$;
D3: black D3;
W11: $11$D black wave; 
$\overline{W11}$: $11$D black wave smeared on $V$;
$11$D BH: light cone M-theory black hole; 
$\overline{11D}$ BH: light cone M-theory black hole smeared on $V$.
}
\label{phase3}
\end{figure}

Figures~\pref{phase2} and~\pref{phase3} depict the phase diagrams of
light cone M-theory on $T^2$ and $T^3$; $V$ here is the radius of the cycles 
(which are chosen to be equal) measured in Planck units. We have similar
observations to the ones made for the previous $1+1$D SYM case. 
In the strong coupling region of SYM on $T^2$, 
the SYM dynamics approaches the
infrared fixed point governing the dynamics of M2 branes --
the conformal field theory dual to M-theory on $AdS_4\times S^7$ 
(in `Poincare' coordinates).
The proper size of the $T^2$ shrinks toward the origin;
at high entropy, the black M2 geometry accurately describes
the low-energy physics, while at low entropy 
the near-horizon geometry is best described
in terms of the IIB theory dual to M-theory on $T^2$~\cite{INCREDIBLE}.
In the $T^3$ case, the diagram reflects the self-duality 
of the D3 branes and M-theory on $T^3$
as reflection symmetry about $V\sim 1$.
The 't Hooft scaling limit 
focusses in on the neighborhood of
the vertical line at $ln\,V/ln\,N\rightarrow\pm\frac13$
(see equation~\pref{geff}).
The structure of all three of these phase diagrams can be checked by
minimizing the Gibbs energies between the various phases identified.

Finally, we conclude with the following observation. 
Starting with a thermodynamic phase in light cone M-theory, 
say for example the lower right corner
phase of the $11$D boosted black hole, using geometrical considerations,
the duality symmetries of M-theory, and the Horowitz-Polchinski
correspondence with the perturbative SYM phase, 
we would be led to conclude that 
light cone M-theory thermodynamics is encoded in the
thermodynamics of SYM QFT. 
Indeed, the Maldacena conjecture asserts that underlying all these phases
is super Yang-Mills theory in various regimes of its parameter space.
Having not known the Matrix conjecture, we would
then have been led to it from Maldacena's proposal. The
Matrix conjecture is a special realization of the
more general statement of Maldacena.
Correspondingly, our ability to discover the low-energy 
theories that yield matrix theory on some background depends
on our ability to understand duality structures with less supersymmetry
in sufficient detail to construct the phase diagram analogous
to figures 1-3.

This introduction and summary set forth all our qualitative results and
conclusions. The computational details can be found in the
next section.

\section{The details}

The theory is parametrized by the Planck scale $\lp$, longitudinal
radius $\r11$, and $p$ circle radii $R$. We define $V\equiv R/\lp$. The
Maldacena or Matrix limit is\footnote{In other words,
$\alpha'\rightarrow 0$, with $\gym^2$ and $R/\ls$ fixed.
Our notational conventions are the same as in \cite{LMS}.}
\bb\label{limit}
\lp\rightarrow 0\mbox{ ,  with  }\lp^2/\r11 \mbox{ and }\lp/R\mbox{ fixed.}
\ee
We will begin by treating
various geometries for $1\le p\le 3$ collectively, with $p$ a variable;
we will then have to analyze separately regions of the phase
diagram where significant differences arise between the different cases.

\subsection{Stretched and smeared geometries}

We first study the phases where the horizon is
stretched or smeared along compact directions. The former case corresponds
to situations where an extended object is wrapped on compact directions;
the geometry cannot localize on such cycles by virtue of the symmetry
structure of the metric.
The latter case corresponds to solutions which are smeared along cycles
because they would otherwise not `fit in the box'; these are prone
to localization transitions to more symmetric, entropically favored 
horizon geometries. 
Both cases are endowed with the isometries of translation along the cycles.
The solutions are parametrized by two harmonic functions
\bb
h=1-\lk(\frac{r_0}{r}\re)^{7-p}\ ,
\ee
\bb
H=1+\lk(\frac{q}{r}\re)^{7-p}\ .
\ee
From the area-entropy relation of the corresponding geometries, we have
\bb
r_0^{9-p}\sim \lk(S^2/N\re) \lp^{9-p} V^{-p}\ ,
\ee
while Gauss's law yields
\bb
q^{7-p}\sim \frac{\lp^{9-p}}{\r11^2}\frac{N}{V^p}\ .
\ee
This last statement is valid for $q\gg r_0$, which is the case in the
limit~\pref{limit} with $S$, $N$ and $V$ finite. For large $N$, when
interested in the thermodynamic limit of 
a large number of degrees of freedom,
this limit is certainly satisfied.

We will track the full form of the geometries; at the end of the day, 
the conjecture requires us to look at the near horizon
region, $q\gg r$.

\sect{Dp brane phase}
This phase is described by the equation of state of black Dp branes
\bb\label{Dpeos}
E\sim \frac{\r11}{\lp^2} \lk( \frac{S^2}{N} \re)^{\frac{7-p}{9-p}}
V^{2\frac{p}{9-p}}\ ,
\ee
and comprises of the geometries of stretched black Dp branes, smeared
D0 branes, and smeared $11$D black waves.
We now analyze each in turn.

The {\it black Dp brane (D1,D2,D3)} is given by the solution~\cite{HORSTROM}
\bb\label{1D1metric}
ds^2=H^{-1/2} \lk( -h dt^2+d{\vec y}_p^2\re) + H^{1/2} \lk( h^{-1} dr^2
+r^2 d\Omega_{8-p}^2\re)\ ,
\ee
\bb
e^\phi=\gs H^{(3-p)/4}\ ,
\ee
\bb
F_{rty}=\gs^{-1} \del_r H^{-1}\ .
\ee
The theory is parametrized by
\bb\label{dualrel}
\alpha'=\frac{\lp^3}{\r11}\quad,\qquad
\gs=\lk(\frac{\r11}{\lp}\re)^{(3-p)/2} V^{-p}\ ,
\ee
and the coordinates $\vec y$ are compactified on 
circles of size $\lp^2 V^{-1}/\r11$.
The relevant restrictions are:
\begin{itemize}
\item Small coupling at the horizon requires
\bb\label{cpl2}
\lk( N^{8-p} S^{p-7}\re)^{3-p} V^{3p(p-7)} < 1\ .
\ee
Otherwise, for $p=1,3$, we have to go 
to the S-dual geometry of black IIB fundamental strings or black D3 branes
respectively; for $p=2$, we need to analyze smeared black M2 branes.

\item Curvature at the horizon smaller than the string scale requires
\bb\label{1corr}
N^{6-p} V^{-3p} > S^{3-p}\ .
\ee
Otherwise, we invoke the principle of correspondence -- 
a perturbative $p+1$D SYM phase emerges.

\item Requiring the cycle size of the $y$'s at the horizon
to be greater than the string scale yields
\bb\label{1D1T}
S>N^{\frac{8-p}{7-p}} V^{3\frac{6-p}{7-p}}\ .
\ee
Otherwise, we go to the T-dual geometry of smeared D0 branes.
\end{itemize}

\noindent
The {\it smeared D0 brane ($\overline{D0}$)} is the T-dual of~\pref{1D1metric} on the 
torus $T^p$ of the $y_i$
\bb\label{1D0metric}
ds^2=-H^{-1/2} h dt^2 + H^{1/2} \lk( dy_p^2 + h^{-1} dr^2
+r^2 d\Omega_{8-p}^2\re)\ ,
\ee
\bb
e^\phi=\gs H^{3/4}\ ,
\ee
\bb
A_t=\gs^{-1} \lk( H^{-1}-1\re)\ .
\ee
The theory is parametrized by
\bb\label{mainrel}
\alpha'=\frac{\lp^3}{\r11}\quad,\qquad
\gs=\lk(\frac{\r11}{\lp}\re)^{3/2} \ ,
\ee
and the new coordinates $\vec y$ are compactified on the scale $\lp V$.
The restrictions are:
\begin{itemize}
\item Small coupling at the horizon requires
\bb\label{newcoupl}
S> N^{\frac{8-p}{7-p}} V^{\frac{p}{p-7}}\ .
\ee
Otherwise, we have to lift to an $11$D M theory black wave solution.

\item The correspondence point is~\pref{1corr}.

\item Requiring the `box' size of the $y_i$
at the horizon to be smaller than the object
yields
\bb\label{1D0loc}
S>V^{9/2} N^{1/2}\ ,
\ee
independent of $p$.
Otherwise, the system collapses into a localized D0 geometry along 
the torus $T^p$ parametrized by the $y_i$.
\end{itemize}

\noindent
The {\it smeared $11$D black wave ($\overline{W11}$)} is the M-lift of~\pref{1D0metric}
\bb\label{1W11}
ds^2=\lk(H-1\re) \lk( dx_{11} -dt\re)^2+dx_{11}^2-dt^2
+H^{-1} \lk(1-h\re) dt^2+dy_p^2+h^{-1} dr^2+r^2d\Omega_{8-p}^2\ ,
\ee
The theory is parametrized by the 
light cone M theory Planck scale $\lp$,
the coordinates $\vec y$ are compactified on $\lp V$ as before,
while $x_{11}$ lives on $\r11$.
The new constraints are:
\begin{itemize}
\item Requiring the size of $x_{11}$ measured at the horizon to be
greater than the Planck scale leads to 
the reverse of~\pref{newcoupl}, patching
back to the smeared D0 geometry.

\item The size of the $y$ cycles measured
at the horizon must be greater than the Planck scale
\bb\label{1W11coupl}
V>1\ .
\ee
Otherwise, we have to dimensionally reduce on a $y$ to a IIA geometry.
For $p=1$, this will be a IIA black wave. For $p=2$, the new geometry
will have cycles smaller than the string scale; so, we need to go to 
the T-dual vacuum representing a IIB black wave; this is of course
just the well known duality between M theory on a shrinking $T^2$ and IIB
on the circle. For $p=3$, we emerge into a dual M-theory with a black wave
geometry.

\item Requiring the `box' size associated with $x_{11}$ measured 
at the horizon to be smaller than the object yields
\bb\label{SNcurve}
S>N\ .
\ee
Otherwise, the system collapses into an $11$D black hole 
smeared along $\vec y$.

\item Requiring the `box' size associated with the $y_i$ measured 
at the horizon to be smaller than the object yields~\pref{1D0loc}; 
the new geometry would be an $11$D black wave localized on
the $y_i$.

\end{itemize}

\sect{Smeared $11$D black hole ($\overline{11D}$ BH)}
This is the Schwarzschild black hole in 
light cone M-theory 
on $T^p$
with Planck scale $\lp$, and torus radii $\r11$ and $R$, 
such that the solution
is smeared on $T^p$. The form of the metric will be discussed later.
The equation of state is given by
\bb\label{1smearedhole}
E\sim \lk( \frac{\r11}{N} \frac{1}{\lp^2} \re) V^{2\frac{p}{9-p}} 
S^{2\frac{8-p}{9-p}}\ .
\ee

\begin{itemize}
\item The correspondence principle yields
\bb
S>V^p\ ,
\ee
which will always be satisfied. Particularly, for $p=0$, we have the
statement $S>1$.
At large $N$, this very low entropy
regime passes out of the region of interest.

\item For $V<1$, we reduce to a IIA geometry; for $p=1$, this is a IIA
hole; for $p=2$, we need to go to the T-dual IIB hole solution
for reasons discussed above;
for $p=3$, we have a smeared $11$D hole in the dual M theory
on $T^3$.

\item Finally,
the localization transition can be found by equating~\pref{1smearedhole}
to the energy of the $p=0$ phase, yielding
\bb\label{v9loc}
S\sim V^9\ .
\ee
\end{itemize}

\subsection{Localized geometries}

The localized solutions are obtained from 
the above geometries~\pref{1D0metric}
and~\pref{1W11} when the box size associated with the $y_i$ 
measured at the horizon
becomes greater than the size of the object. The system
collapses then into a more symmetric, entropically favored
solution by the substitution
\bb
d{\vec y}_p^2+h^{-1} dr^2+r^2 d\Omega_{8-p}^2\rightarrow
h^{-1} dr^2 + r^2 d\Omega_8\ .
\ee
The entropy-area relationship changes to
\bb\label{loc1}
r_0\sim \lp S^{2/9} N^{-1/9}\ ,
\ee
and the functions $H$ and $h$ become
\bb\label{loc2}
h\rightarrow 1-\lk(\frac{r_0}{r}\re)^7\ ,
\ee
\bb\label{loc3}
H\rightarrow 1+c \lk(\frac{\lp^9}{\r11^2} N \re) \frac{1}{r^7}\ .
\ee
Here $c$ is some numerical constant.

In the subsequent subsections, parameters and solutions can be obtained
from their smeared relatives, with the changes just described.

\sect{D0 phase}
The equation of state is given by~\pref{Dpeos} with $p=0$
\bb\label{D0eos}
E\sim \lk(\frac{\r11}{N} \frac{1}{\lp^2}\re) S^{14/9} N^{2/9}\ ,
\ee
and consists of two patches, a localized black D0 and
a localized $11$D black wave.

The restrictions on the {\it localized black D0 (D0)} are:

\begin{itemize}

\item Coupling near the horizon must be small
\bb
S>N^{8/7}\ .
\ee
Otherwise, we lift to the localized $11$D black wave solution.

\item The curvature near the horizon must be small with respect to the string
scale
\bb\label{1FTlocal}
S<N^2\ .
\ee
Otherwise, by the correspondence principle,
we go to the perturbative
$p+1$D SYM phase. From the SYM point of view, this point is where the 
thermal wavelength becomes the order of the box size (dual to $R$); the
perturbative excitations are frozen on the circle; this then naturally
maps onto the black D0 geometry.

\end{itemize}

\noindent
The {\it localized $11$D black wave (W11)} 
sews onto the previous one when the size of $x_{11}$ at the
horizon is the Planck scale, and localizes on this cycle unless
\bb\label{D0end}
S>N\ .
\ee
Beyond this point, the vacuum is that of a 
light cone $11$D Schwarzschild
black hole which we discuss next.

\sect{$11$D light cone black hole (11D BH)}
Let us discuss in some generality the light cone black hole.
This is a localized Schwarzschild solution with momentum along $x_{11}$.
Using the Polchinski-Seiberg 
procedure~\cite{WHYSEIB,HPOLCH}, a metric of the form
\bb
ds^2=g_{00} dt^2 + g_{11} dx_{11}^2+\cdots\ .
\ee
is put in the light cone frame by a large boost
\bb
e^\alpha\sim \frac{N}{M \r11}\ ,
\ee
where $M$ is the rest mass of the original solution, and $\r11\rightarrow 0$
in the Matrix/Maldacena limit~\pref{limit}. The metric becomes
\bb\label{LCmetric}
ds^2=\frac{1}{2} (g_{11}-g_{00}) dx_+ dx_-
+ \frac{N^2}{M^2 \r11^2} (g_{00} + g_{11}) dx_-^2\ .
\ee
This can further be mapped onto the DLCQ frame by infinite
boost $e^\delta \sim R_+/\r11$, with $R_+$ finite,
multiplying the $dx_-$ by $\r11/R_+$. We will not worry here about this
additional mapping, and keep statements in the light cone language, 
maintaining $\r11$ in the limit~\pref{limit}.

The black hole solution in $d+3$ dimensions is
\bb
ds^2=-f dt^2+f^{-1} d{\tilde{r}}^2 +{\tilde{r}}^2 d\Omega_{d+1}^2\ ,
\ee
with $f \equiv 1-\lk(\frac{r_0}{\tilde{r}}\re)^d$.
To boost it, we first choose isotropic coordinates; this can be achieved by
the coordinate transformation
$ {\tilde{r}}^d = (\rho^d/4) ( 1+(r_0/\rho)^d )^2$, 
with $\rho^2=x_{11}^2+r^2$.
The metric becomes
\bb
ds^2=-\lk[\frac{1-(r_0/\rho)^d}{1+(r_0/\rho)^d}\re]^2 dt^2
+ \lk(1+\lk(\frac{r_0}{\rho}\re)^d\re)^{4/d} (dx_{11}^2
+dr^2+r^2d\Omega^2)\ .
\ee
Now boost on $x_{11}$ \`{a} la Polchinski-Seiberg, compactify on $R_{11}\rightarrow 0$;
the metric becomes of the form~\pref{LCmetric}, with $1\pm(r_0/\rho)^d$
replaces by
\bb
1\pm \sum_n \frac{r_0^d}{(r^2 + (N^2/M^2R_{11}^2)
(x_-+2\pi n R_{11})^2)^{d/2}}\ .
\ee

In our case, $d=8$ for an $11$D light cone black hole. The smeared
light cone hole encountered above, and other
light cone holes we will encounter, can be obtained from this geometry by
smearing on cycles, and duality transformations.
The equation of state is given by~\pref{1smearedhole} with $p=0$
\bb\label{11Dhole}
E\sim \lk(\frac{\r11}{N} \frac{1}{\lp^2}\re) S^{16/9}\ .
\ee
Minimizing this energy with respect to the one corresponding to
the light cone hole smeared on $R$ (equation~\pref{1smearedhole}
with $p=1,2,3$), 
one finds the transition curve~\pref{v9loc}.

\subsection{Perturbative $p+1$D SYM}

Here, weakly coupled gluonic excitations dominate the dynamics.
The scaling of the equation of state is obtained by dimensional analysis
\bb
E\sim \lk(\frac{\r11}{N} \frac{1}{\lp^2}\re) V N^{\frac{p-2}{p}}
S^{\frac{p+1}{p}}\ .
\ee
This regime sews onto the localized D0 brane solution 
by the correspondence principle at~\pref{1FTlocal}. 
This can also be checked by setting the thermal wavelength
equal to the dual box size in the perturbative field theory, 
or by minimizing the Gibbs energies between the
localized D0 and perturbative SYM phases.
For large Yang-Mills couplings, this SYM phase sews onto
the Dp brane geometry at~\pref{1corr}; this is again an application of
the correspondence principle.

\subsection{Comments on correspondence curves}

In~\cite{RABIN}, an additional 
phase labeled Super Quantum Mechanics was identified on the 
temperature -- $g_s N$ plane. On the $\ln S$-$\ln V$ plane, this 
corresponds to the single line segment at $S\sim N^2$
separating the perturbative SYM phase from the
black $D0$-brane phase in figures 1--3.  Two critical
phenomena are identified with the same line. 
From the higher entropy side, the perturbative SYM freezes its
dynamics on the torus at temperatures of order 
$T_c^{(1)}\sim \r11 V/\lp^2\sim \Sigma^{-1}$; 
using the perturbative SYM equation of state, this corresponds to $S\sim N^2$.
From the side of lower entropies, 
the correspondence point \pref{1FTlocal} occurs again at $S\sim N^2$;
using the D0 phase's equation of state \pref{D0eos}, 
this corresponds to temperatures
of order $T_c^{(2)}\sim \r11 N^{1/3} / \lp^2$. 
On this line, there is a phase
whose entropy remains constant while the temperature changes; \ie\ the
specific heat vanishes. There may be interesting physics to be studied here
via SYM dynamics.

The two sides of the correspondence curve are both statements about
the effective SYM coupling becoming of order one.
Equation~\pref{geff} applies for all $p$:\footnote{Ideas relating to
the scaling of the effective coupling
for $p=0$ were also discussed in~\cite{MEANA}.}
\bb\label{geff2}
g_{\rm eff}^2\sim V^{-p} N \lk(\frac{T \lp^2}{\r11}\re)^{p-3}\ .
\ee
We can use this also for D0 branes on the dual torus; 
T-duality on the Dp branes
is encoded in this relation, as can be seen 
by comparing equations~\pref{dualrel}
and \pref{mainrel}. 
The resulting SYM zero mode Lagrangian has a coupling
$\gym^2/\mbox{Vol}\sim \gym^2 (\r11 V/\lp^2)^p$. Note also that we
use the temperature $T\sim E/S$, not the energy\footnote{The notation
in~\cite{PEETPOLCH} is such that $E$ represented temperature $T$.
At finite temperature, 
the energy scale relevant to the dynamics is set naturally by the 
temperature. For supergravity probes with thermodynamic character,
the UV-IR correspondence relates the
field theory temperature to extent in the bulk; this is the same
as identifying area in the bulk with entropy in the 
field theory.}.
Using the equation of state of the perturbative SYM, we
translate the statement $g_{\rm eff}^2\sim 1$ with $p\neq 0$
to equation~\pref{1corr}. Using equation~\pref{Dpeos}, we find the
equipotentials of the effective coupling in the Dp phase
\bb
g_{\rm eff}^2\sim \lk( N^{6-p} S^{p-3} V^{-3p}\re)^{\frac{5-p}{9-p}}\ .
\ee
For $p<4$ and in the Dp phase domain, 
the effective coupling increases diagonally on the diagrams
as we move toward lower entropies and smaller volumes $V^p$.
Using the equation of state of the localized D0 phase and
~\pref{geff2} with $p=0$, we obtain~\pref{1FTlocal} for 
$g_{\rm eff}^2\sim 1$.
The equipotentials change in the D0 phase for all three diagrams
\bb
g_{\rm eff}^2\sim \lk( \frac{N^2}{S}\re)^{5/3}\ .
\ee
The coupling increases from one at $S\sim N^2$ as we lower the entropy toward
the $11$D black hole phase.
From SYM physics, both correspondence curves are where the effective 
coupling is of order one; the localization effect at $S\sim N^2$ changes
this effective coupling appropriately. 
It is tempting to generalize this observation and propose that
the effective coupling in the field theory is to be always identified
with curvature scale in string units in the supergravity. This becomes a
a non-trivial statement about the effective degrees of freedom and the
dynamics of the field theory in its non-perturbative regimes.

\subsection{Specialized treatments: $p=1$}

\sect{Black IIA wave and the black IIB string}
This region of the
phase diagram is characterized by the 
equation of state~\pref{Dpeos} with $p=1$;
it comprises of two geometries patched against the D1 brane and M wave
phases studied above.

The {\it IIA black wave (W10)} is the dimensional reduction of the 
previous geometry~\pref{1W11} on $y$
\bb\label{1IIAwave}
ds^2=\lk(H-1\re) \lk( dx_{11} -dt\re)^2+dx_{11}^2-dt^2
+H^{-1} \lk(1-h\re) dt^2+h^{-1} dr^2+r^2d\Omega_7^2\ ,
\ee
The theory is parametrized by
\bb
\alpha'\sim \lp^2 V^{-1}\ ,
\ee
\bb
\gs\sim V^{3/2}\ .
\ee
The coordinate $x_{11}$ is compactified on $\r11$.
The two relevant restrictions are:
\begin{itemize}
\item The size of the cycle $x_{11}$ measured
at the horizon must be smaller than the size of the object
\bb\label{1r11loc}
S>N\ .
\ee
Otherwise, the system collapses into the smeared 
$11$D black hole; strictly speaking, its
dimensional reduction on the smeared direction $y$.

\item The size of $x_{11}$ at the horizon being greater than the string
scale is the statement
\bb
S<V^{1/2} N^{7/6}\ .
\ee
Otherwise, we go to the T-dual geometry of black IIB fundamental strings.
\end{itemize}

\noindent
The {\it black IIB string (F1)} is the T-dual of~\pref{1IIAwave} on $x_{11}$
\bb\label{1IIBstring}
ds^2=H^{-1}\lk(dx_{11}^2-h dt^2\re) +h^{-1} dr^2+r^2\Omega_7^2\ ,
\ee
\bb
e^\phi=\gs H^{-1/2}\ ,
\ee
\bb
B_{11,t}=H^{-1}-1\ .
\ee
The theory is parametrized by
\bb
\alpha'=\lp^2 V^{-1}\quad,\qquad
\gs=\frac{\lp}{\r11} V \ ,
\ee
and the new coordinate $x_{11}$ is compactified on
$\lp^2 V^{-1}/\r11$.
The two new constraints are:
\begin{itemize}
\item Curvature at the horizon must be smaller than the string scale
\bb\label{1MS1}
S>V^{-3/2} N^{1/2}\ .
\ee
Beyond this point, we call upon the correspondence principle and identify 
a new phase consisting of a Matrix string; more on this phase later.

\item For strong couplings, we patch to the D1 geometry through S-duality
at~\pref{cpl2}.
\end{itemize}

\sect{$10$D IIA black hole (10D BH)}
The {\it $10$D IIA black hole}
is obtained by dimensional reduction of the smeared $11$D solution.
Its equation of state is given by~\pref{1smearedhole}.
Its correspondence point occurs at
\bb\label{1MS2}
S\sim V^{-3}\ .
\ee
The Matrix string phase emerges beyond this point.

\sect{Matrix string phase}
Correspondence curves delineate the fundamental string and smeared black hole
geometries identified above. Both of these transition curves~\pref{1MS1} 
and \pref{1MS2} are accounted for by
Gibbs energy minimization with respect
to the equation of state of a Matrix string
\bb
E\sim \lk(\frac{\r11}{N} \frac{1}{\lp^2}\re) V S^2\ .
\ee
We conclude that the new phase beyond these geometries is that of a
Matrix string; \ie\ a $Z_N$ holonomy is induced at strong Yang-Mills
coupling that sews the D-strings into a coil or `slinky'. 
This physics is then associated
with the emergence of new order and symmetry.

\subsection{Specialized treatements: $p=2$}

\sect{The IIB black wave and smeared black membranes}
The equation of state is given by~\pref{Dpeos}; this region comprises of
two patches.

The {\it IIB black wave ($\overline{WB}$)} 
is the dimensional reduction of~\pref{1W11}
on one of the $y$'s, and a further T dualization on 
the other; the
latter step is needed because, by virtue of focusing on a square torus,
the intermediate IIA theory lives on a circle smaller than its string scale.
As mentioned earlier, this is the well known M-IIB duality. 
The vacuum is given by
\bb\label{2IIBwave}
ds^2=\lk(H-1\re) \lk( dx_{11} -dt\re)^2+dx_{11}^2-dt^2
+H^{-1} \lk(1-h\re) dt^2+dz^2+h^{-1} dr^2+r^2d\Omega_6^2\ ,
\ee
The theory is parametrized by
\bb
\alpha'\sim \lp^2 V^{-1}\quad,\qquad
\gs\sim \frac{R}{R} \equiv 1\ .
\ee
The coordinate $x_{11}$ is compactified on $\r11$, while $z$ lives on
$\lp V^{-2}$.
We note that we are at the self-dual point of the IIB theory. By the
S-duality symmetry of the low energy effective dynamics, 
the structure of the geometry is valid at this point.
The new restrictions are:
\begin{itemize}
\item Cycle $x_{11}$ measured
at the horizon must be greater than the string scale 
\bb\label{2D2TtoIIB}
S<V^{3/10} N^{6/5}\ .
\ee
Otherwise, we have to go to a IIA geometry by T-duality.
This will be a black fundamental string. Its coupling at the horizon is found
to be large in this regime; we therefore have to also lift to 
an M theory, yielding black smeared membranes.

\item The size of $x_{11}$ measured
at the horizon must be smaller than the size of the object
\bb
S>N\ .
\ee
Otherwise, the system collapses into an $11$D light cone
black hole smeared on the $R$'s; more accurately, the emerging phase
is the M-IIB dual of such a hole, for the same reasons discussed above.

\item The cycle size of $z$ measured
at the horizon must be smaller than the size of the object
\bb
S>V^{-6} N^{1/2}\ .
\ee
Otherwise, we localize the IIB black wave on the circle $z$.

\end{itemize}

\noindent
As mentioned above, we encounter a 
{\it smeared black membrane ($\overline{M2}$)} phase by a chain
of two dualities; a T-duality of the previous solution, and then a lift
to an M theory. The geometry becomes
\bb\label{2M2geom}
ds^2=H^{-2/3}\lk(dx_2^2-h dt^2\re) +
H^{1/3} \lk( dz^2+h^{-1} dr^2+r^2 d\Omega_6^2\re)\ ,
\ee
\bb
A_{12t}=1-H^{-1}\ .
\ee
The theory is parametrized by the Planck scale
\bb
\tilde{l}_{{\rm Pl}}^3=\frac{\lp^4}{\r11} V^{-2}\ ,
\ee
the coordinates $x_{1,2}$ are compactified on $\lp V^{-2}$
and the new $z$ coordinate lives on $\lp^2 V^{-1} /\r11$.
The only new constraint is:
\begin{itemize}
\item Requiring the `box' size of $z$ measured
at the horizon to be smaller than the object yields
\bb
S>V^{-6} N^{1/2}\ .
\ee
Otherwise, we need to analyze a black localized membrane geometry.

\end{itemize}

\sect{Smeared IIB black hole ($\overline{10D}$ BH)}
This phase descends by localization from the IIB wave encountered above,
or by dimensional reduction/T-duality from the $11$D smeared light cone hole.
The equation of state is given by~\pref{1smearedhole}. It will undergo
localization on the IIB circle to a delocalized $10$D hole; the transition
point is analyzed below.

\sect{Localizations}
The membrane, IIB wave and IIB hole encountered above are all subject
to localization transitions on the dual circle; this corresponds to
transtions between $7$ and $8$ noncompact
transverse space dimensions. 
The metrics~\pref{2IIBwave} and~\pref{2M2geom} undergo the substitution
$dz^2+h^{-1} dr^2 +r^2 d\Omega_6^2\rightarrow
h^{-1} dr^2 +r^2 d\Omega_7^2$, and
\bb
h\rightarrow 1-\lk(\frac{r_0}{r}\re)^6\ ,
\ee
\bb
H\rightarrow 1-c \lk(\frac{\lp^8 N V^{-4}}{\r11^2}\re)\lk(\frac{1}{r^6}\re)\ ,
\ee
\bb
r_0\sim \lp N^{-1/8} V^{-1/2} S^{1/4}\ .
\ee
We next discuss these three localized phases.

\sect{Membrane phase}
This phase consists of two patches.
There is the {\it localized phase of black membranes (M2)};
its equation of state is
\bb
E\sim \lk(\frac{\r11}{N}\frac{1}{\lp^2} \re) V S^{3/2} N^{1/4}\ .
\ee
It is restricted by

\begin{itemize}

\item The correspondence point
\bb
N>1\ .
\ee

\item The requirement that the cycles on which it is wrapped are bigger
than the Planck scale
\bb\label{2memwave}
S>N^{7/6}\ .
\ee
We then lift to the localized black IIB wave.

\end{itemize}

\noindent
The {\it localized black IIB wave (WB)} patchs onto this
membrane phase at~\pref{2memwave} when $x_{11}$ is the string scale, \ie\ the
T-duality-M lift point.  It localizes on $\r11$ unless $S>N$. Below
this entropy, we have a fully localized IIB black hole.

\sect{IIB black hole (10D BH)}
This is the IIB hole localized on the dual IIB radius. It descends by
$\r11$ localization from the previous geometry; its equation of state is
\bb
E\sim \lk(\frac{\r11}{N}\frac{1}{\lp^2} \re) V S^{7/4}\ .
\ee
Minimizing its Gibbs energy with respect to that of the smeared $11$D
hole~\pref{1smearedhole} yields the transition curve
\bb
S\sim V^{-12}\ .
\ee

\subsection{Specialized treatements: $p=3$}

For $p=3$, both the IIB S-duality of the D3 brane and the M theory
duality on $T^3$ correspond to the transition curve at $V\sim 1$. The
phase diagram has a reflection symmetry about $V\sim 1$ and no new
phases arise. 

\subsection{The final picture}

Summarizing our analysis, we map out the thermodynamic phase
diagram of M theory on $T^p$, or $p+1$D SYM on the circle 
(Figures~\ref{phase1}-\ref{phase3}). 
Using the various equations
of state, minimizing the Gibbs energies with respect to each
other, one finds the scaling properties of all 
the transition curves identified 
through geometrical considerations.
Finally, we conclude by identifying the interesting self-dual point at
\bb
V\sim 1\quad ,\qquad S\sim N^{\frac{8-p}{7-p}}\ .
\ee

\vskip 2cm
\noindent
{\Large{{\bf Acknowledgments}}}:
This work was supported by DOE grant DE-FG02-90ER-40560.

\newpage

\providecommand{\href}[2]{#2}\begingroup\raggedright\endgroup

\end{document}